\documentclass[final,5p,times,twocolumn,nopreprintline]{elsarticle}
\usepackage{amsmath,slashed,booktabs,dsfont}
\usepackage{graphicx,graphics}
\usepackage{dcolumn}
\usepackage[hyperfootnotes=false]{hyperref}
\usepackage{xspace}
\usepackage{color}
\usepackage{balance}
\usepackage{multirow}
\usepackage{multicol}

\usepackage{fancyhdr}
\fancypagestyle{firstpage}{%
	
	\lhead{}
	\rhead{\small PSI-PR-23-03, ZU-TH 08/23}
}
\newcommand{\GeV}{\,\text{GeV}}

\newcommand{\qvec}{\mathbf{q}}
\newcommand{\kvec}{\mathbf{k}}
\renewcommand{\Im}{\text{Im}\,}
\renewcommand{\Re}{\text{Re}\,}

\newcommand{\beq}{\begin{equation}}
\newcommand{\eeq}{\end{equation}}
\newcommand{\Lagr}{\mathcal L}
\newcommand{\mN}{m_N}
\newcommand{\MP}{M_P}

\allowdisplaybreaks[4]

\begin{document}

\renewcommand{\theequation}{\arabic{equation}}

\begin{frontmatter}
 
\title{Rescattering effects in nucleon-to-meson form factors\\[1mm] and application to tau-lepton-induced proton decay}

\author[PSI,Zurich]{Andreas Crivellin}
\author[Bern]{Martin Hoferichter}

\address[PSI]{Paul Scherrer Institut, 5232 Villigen PSI, Switzerland}
\address[Zurich]{Physik-Institut, Universit\"at Z\"urich, Winterthurerstrasse 190, 8057 Z\"urich, Switzerland}
\address[Bern]{Albert Einstein Center for Fundamental Physics, Institute for Theoretical Physics, University of Bern, Sidlerstrasse 5, 3012 Bern, Switzerland}

\begin{abstract}
 Nucleon decays put extremely stringent bounds on baryon-number-violating interactions. However, in case the corresponding operators involve only $\tau$ leptons, the direct two-body decays, e.g., $p\to\pi^0 \tau^+$, are kinematically not allowed and nucleon decay can only proceed via an off-shell $\tau$, leading to $p\to \pi^0\ell^+\nu_\ell\bar\nu_\tau$. To calculate such processes, the momentum dependence of the form factors for nucleon-to-meson transitions, which describe the hadronization of the underlying process at the quark level, is needed. In this work, we point out new isospin and Fierz relations among such proton--kaon matrix elements and calculate the momentum dependence of the nucleon-to-meson form factors from the universal final-state interactions in terms of pion--nucleon or kaon--nucleon scattering phase shifts. We use these results to derive novel limits on the Wilson coefficients of the baryon-number-violating dimension-$6$ operators involving a $\tau$ lepton, which were previously unconstrained.        
\end{abstract}

\end{frontmatter}

\thispagestyle{firstpage}

\section{Introduction}

\nocite{Pati:1973uk,Georgi:1974sy,Fritzsch:1974nn}

\begin{table}[t]
	\renewcommand{\arraystretch}{1.3}
	\centering
	\scalebox{0.95}{
	\begin{tabular}{l r r}
		\toprule
		 Channel & Limit [$10^{30}$ y] & Reference \\\midrule
		 $p\to \pi^0 e^+$ & $2.4\times 10^{4}$ & \cite{Super-Kamiokande:2020wjk}\\
		 $p\to \pi^0 \mu^+$ & $1.6\times 10^{4}$ & \cite{Super-Kamiokande:2020wjk}\\
		 $p\to \pi^+ \bar \nu$ & $3.9\times 10^{2}$ & \cite{Super-Kamiokande:2013rwg}\\
		 $p\to K^0e^+$& $1.0\times 10^3$&\cite{Super-Kamiokande:2005lev}\\
		 $p\to K^0\mu^+$& $3.6\times 10^3$&\cite{Super-Kamiokande:2022egr}\\
		 $p\to K^+\bar\nu$& $5.9\times 10^3$&\cite{Super-Kamiokande:2014otb}\\
		  $p\to \eta e^+$& $1.0\times 10^4$&\cite{Super-Kamiokande:2017gev}\\
		 $p\to\eta\mu^+$& $4.7\times 10^3$&\cite{Super-Kamiokande:2017gev}\\\hline
		 $n\to \pi^- e^+$ & $5.3\times 10^3$ & \cite{Super-Kamiokande:2017gev}\\
		 $n\to \pi^- \mu^+$ & $3.5\times 10^3$ & \cite{Super-Kamiokande:2017gev}\\
		 $n\to \pi^0 \bar \nu$ & $1.1\times 10^{3}$ & \cite{Super-Kamiokande:2013rwg}\\
		 $n\to K^0\bar\nu$& $1.3\times 10^2$&\cite{Super-Kamiokande:2005lev}\\
		 $n\to \eta\bar\nu$& $1.6\times 10^2$&\cite{McGrew:1999nd}\\\midrule
		 $p,n\to e^+ X$ & $0.6$ & \cite{Learned:1979gp}\\
		 $p,n\to \mu^+ X$ & $12$ & \cite{Cherry:1981uq}\\
		\bottomrule
	\end{tabular}
	}
	\caption{Limits on nucleon lifetimes for various decay channels (all at $90\%$ C.L.)~\cite{ParticleDataGroup:2022pth}.}  
	\label{tab:lifetime}
\end{table}

The potential simultaneous violation of baryon and lepton number is one of the frontiers of searches for physics beyond the Standard Model (BSM). Motivated by the prediction of proton decay in grand unified theories~\cite{Pati:1973uk,Georgi:1974sy,Fritzsch:1974nn}, limits on the nucleon decay width for various modes have been set that for some channels exceed $10^{34}$ years, see Table~\ref{tab:lifetime}. In the future, Hyper-Kamiokande~\cite{Hyper-Kamiokande:2018ofw} will be able to improve these limits by another order of magnitude, projecting  $7.8\times 10^{34}$ years for $p\to \pi^0 e^+$ and $3.2\times 10^{34}$ years for $p\to K^+\bar \nu$ (at $90\%$ C.L.),  while competitive new limits, especially for kaon modes, are also expected from DUNE~\cite{DUNE:2020ypp} and JUNO~\cite{JUNO:2021vlw}. Most importantly, these improvements not only pertain to the most sensitive two-particle channels, but also for generic searches such as $p\to e^+ X$ major advances are projected.    

One can interpret such limits model independently in terms of effective operators~\cite{Wilczek:1979hc,Weinberg:1979sa}, which are a subset of the full SM effective field theory (SMEFT) Lagrangian~\cite{Buchmuller:1985jz,Grzadkowski:2010es}. The resulting constraints from direct two-body decays such as $p\to \pi^0\ell$ are in general extremely stringent and push the BSM scale to at least $10^{14}\GeV$ in case of tree-level effects with order-one couplings. However, in case only $\tau$ leptons are involved, the decay of a nucleon is kinematically not allowed, so that the related SMEFT coefficients have not been constrained so far.\footnote{Note that because lepton flavor is exactly conserved in the SM with massless neutrinos, and only marginally violated by the light active neutrino masses, the renormalization group (RG) evolution from the BSM scale to the nucleon scale does not generate operators involving muons or electrons if these coupling are absent at the matching scale. Therefore, contrary to the case with heavy quarks~\cite{Helo:2019yqp}, these operators do not contribute to proton decay even at the loop level.}

Instead, since nucleon decay can be mediated by an off-shell $\tau$, processes such as $p\to P\ell^+\nu_{\ell}\bar\nu_\tau$ with $\ell=e,\mu$ and $P=\pi,K,\eta$ do probe these couplings. They can be constrained by the inclusive limits given in the last panel of Table~\ref{tab:lifetime}, but require knowledge of the momentum dependence of the hadronic part of the matrix elements. In fact, the value of such inclusive limits as generic tests of baryon number violation has already been stressed in Ref.~\cite{Heeck:2019kgr}. In either case, to constrain the underlying Wilson coefficients at the partonic level, the hadronization of the effective operators must be calculated. Some relations among these nucleon-to-meson from factors can be derived using chiral perturbation theory (ChPT)~\cite{Claudson:1981gh,Chadha:1983sj,Chadha:1983mh,Kaymakcalan:1983uc,JLQCD:1999dld}, but especially estimates of their normalization have remained model dependent for a long time~\cite{Gavela:1981cf,Okazaki:1982eh,Martin:2011nd}. In the last years, calculations in lattice QCD~\cite{Hara:1986hk,JLQCD:1999dld,CP-PACS:2004wqk,Aoki:2013yxa,Aoki:2017puj,Yoo:2021gql} have matured to the extent that these normalizations have now been reliably determined for many operators~\cite{Yoo:2021gql}, thus rendering limits derived from two-body decays such as $p\to \pi^0 e^+$ much more robust. 

In this work, we focus on the momentum dependence of the form factors as required for the interpretation of the four-body decays sensitive to $\tau$-mediated proton decay.  
As a first step, we obtain new relations among the proton-to-kaon matrix elements from isospin symmetry and Fierz identities. Next, we analyze the unitarity relations that are produced by meson--nucleon rescattering effects, and show that the corresponding constraints allow one to calculate the momentum dependence from known meson--nucleon phase shifts. This strategy generalizes a similar approach for baryon-number-conserving nucleon form factors, in which case rescattering corrections in the two-meson system are resummed~\cite{Hohler:1976ax,Gasser:1990ap,Hoferichter:2012wf,Hoferichter:2015dsa,Hoferichter:2016duk,Cirigliano:2017tqn,Hoferichter:2018zwu}. Finally, we apply these results to derive novel limits on the Wilson coefficients of the operators involving $\tau$ leptons from four-body decays. 

\section{Operators and matrix elements}

As baryon number violation is experimentally extremely well constrained, it is reasonable to assume that it originates from a very high BSM scale. At this scale electroweak symmetry is unbroken and one can use SMEFT~\cite{Buchmuller:1985jz,Grzadkowski:2010es} to reduce the number of possible $B$-violating operators to four, see~\ref{sec:SMEFT}. After electroweak symmetry breaking, the decomposition of the lepton doublets results in the operators
\begin{align}
\label{QABi}
 Q^{AB}_{udu}&=\left[\bar u^c P_A d\right]P_B u,&
 Q^{AB}_{udd}&=\left[\bar u^c P_A d\right]P_B d,\notag\\
 Q^{AB}_{usu}&=\left[\bar u^c P_A s\right]P_B u,&
 Q^{AB}_{usd}&=\left[\bar u^c P_A s\right]P_B d,\notag\\
 Q^{AB}_{dsu}&=\left[\bar d^c P_A s\right]P_B u,&
 Q^{AB}_{uds}&=\left[\bar u^c P_A d\right]P_B s,
\end{align}
whose matrix elements are calculated within lattice QCD~\cite{Yoo:2021gql}. 
In all cases, the contraction of the color indices with a Levi-Civita tensor is implied, the charge-conjugated fields are denoted by the superscript $c$, and projectors $A,B=L,R$ are defined by $P_{L/R}=(\mathds{1}\mp\gamma_5)/2$. The corresponding Wilson coefficients are defined via ${\mathcal L_\ell}= C^{AB}_{\ell_t \alpha}\bar\ell^c_tQ^{AB}_{\alpha}$, ${\mathcal L_\nu}= C^{AB}_{\nu_t \alpha}\bar\nu^c_tQ^{AB}_{\alpha}$, depending on whether the charge of $\alpha=\{udu,\ldots\}$ requires a charged lepton or a neutrino.     

\begin{table}[t]
	\renewcommand{\arraystretch}{1.3}
	\centering
	\scalebox{0.95}{
	\begin{tabular}{l | r r r r}
		\toprule
		$X_i$ & $W_{0}^{X_{iL}}(0)$ & $W_1^{X_{iL}}(0)$ & $W_0^{X_{iR}}(0)$ & $W_1^{X_{iR}}(0)$\\\midrule
		 $U_{1}$ & $0.151(31)$ & $-0.134(18)$ & $-0.159(35)$ & $0.169(37)$\\
		 $S_{1}$ & $0.043(4)$ & $0.028(7)$ & $0.085(12)$ & $-0.026(4)$\\
		 $S_{2}$ & $0.028(4)$ & $-0.049(7)$ & $-0.040(6)$ & $0.053(7)$\\
		 $S_{3}$ & $0.101(11)$ & $-0.075(13)$ & $-0.109(19)$ & $0.080(17)$\\
		 $S_{4}$ & $-0.072(8)$ & $0.024(6)$ & $-0.044(5)$ & $-0.026(6)$\\\midrule
		 $S_{1+2+4}$ & $0.000(0)$ & $0.000(0)$ & $0.000(0)$ & $0.000(0)$\\
		 $S_{2-3-4}$ & $0.000(0)$ & $0.000(0)$ & $0.112(15)$ & $0.000(12)$\\
		\bottomrule
	\end{tabular}
	}
	\caption{Nucleon-to-meson form-factor normalizations from Ref.~\cite{Yoo:2021gql}, in units of $\GeV^{2}$ at $\overline{\text{MS}}$ scale $\mu=2\GeV$. All uncertainties have been combined in quadrature. Reference~\cite{Yoo:2021gql} also quotes results at $q^2=m_\mu^2$, but the observed variation compared to $q^2=0$ is negligible within uncertainties. Results for the linear combinations from Eq.~\eqref{isospinfierz} are given in the last two lines~\cite{Syritsyn:2023}.  $S_{5A}$ was not considered in Ref.~\cite{Yoo:2021gql} due to disconnected diagrams. Note that due to parity invariance $X_{iL}\equiv X_i^{LL}=X_i^{RR}$, $X_{iR}\equiv X_i^{RL}=X_i^{LR}$.}  
	\label{tab:lattice}
\end{table}

Following the notation from Ref.~\cite{Yoo:2021gql}, we label the matrix elements relevant for proton decay as
\begin{align}
 \langle \pi^0|\left[\bar u^c P_A d\right]u_B|p\rangle&=\frac{1}{\sqrt{2}} \langle \pi^+|\left[\bar u^c P_A d\right]d_B|p\rangle\equiv \frac{1}{\sqrt{2}} U_{1}^{AB},\notag\\
 \langle K^0|\left[\bar u^c P_A s\right]u_B|p\rangle&\equiv S_{1}^{AB}, \qquad \langle K^+|\left[\bar u^c P_A s\right]d_B|p\rangle\equiv S_{2}^{AB},\notag\\
 \langle K^+|\left[\bar u^c P_A d\right]s_B|p\rangle&\equiv S_{3}^{AB},\qquad
  \langle K^+|\left[\bar d^c P_A s\right]u_B|p\rangle\equiv S_{4}^{AB},\notag\\
  \langle \eta|\left[\bar u^c P_A d\right]u_B|p\rangle&\equiv S_{5}^{AB},
  \label{US}
\end{align}
where the first relation is known to follow from isospin symmetry~\cite{Aoki:2013yxa}, i.e., applies in the limit of equal up and down quark masses (isospin symmetry also determines the corresponding neutron matrix elements, see~\ref{app:isospin}). 
Since two of the four chirality combinations are redundant due to parity invariance of the strong interactions, $X_i^{LL}=X_i^{RR}$ and $X_i^{RL}=X_i^{LR}$, we follow Ref.~\cite{Yoo:2021gql} and fix the second index to $B=L$, defining $X_{iA}\equiv X_i^{AL}$, $X=U,S$.  
In addition to the known isospin relation, we find that also the kaon matrix elements are not all independent: 
\begin{align}
\label{isospinfierz}
S_{1A}+S_{2A}+S_{4A}&=0,\notag\\
S_{2L}-S_{3L}-S_{4L}&=0,
\end{align}
where the first equation follows from isospin symmetry and the second from Fierz identities. We have checked that these novel relations are indeed fulfilled both by the tree-level ChPT amplitudes from Ref.~\cite{JLQCD:1999dld} and the most recent set of lattice-QCD results~\cite{Yoo:2021gql,Syritsyn:2023}.

\section{Rescattering corrections}

\begin{figure}[t]
\centering
 \includegraphics[width=0.4\linewidth]{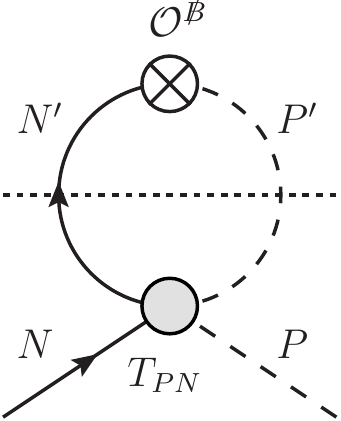}
 \caption{Unitarity diagram for the proton-decay matrix elements. The solid and dashed lines denote nucleons ($N$, $N'$) and pseudoscalar mesons ($P$, $P'$), respectively, while the cross refers to an insertion of the $B$-violating operator ${\mathcal O}^{\slashed{B}}$ and the gray blob to the meson--nucleon scattering amplitude $T_{PN}$. The short-dashed line indicates the cut that produces the imaginary parts in $W_{0,1}(s)$.}
 \label{fig:unitarity}
\end{figure}

By virtue of Lorentz symmetry and parity, the hadronic matrix elements in Eq.~\eqref{US}, of the $B$-violating operators $Q^{AB}_{\alpha}$ defined in Eq.~\eqref{QABi}, between a pseudoscalar meson $P$ with momentum $p'$ and a nucleon $N$ with momentum $p$, can be decomposed as
\beq
X^{AB}_i=P_B\Big[W_{0}^{X_i^{AB}}(s)+\frac{\slashed{q}}{\mN} W_{1}^{X_i^{AB}}(s)\Big] u_N(p),
\eeq
where the two form factors $W_{0,1}^{X_i^{AB}}$ depend on the momentum transfer $s=q^2=(p-p')^2$. Since the strong final-state interactions are universal, depending only on the isospin in a given channel, we will drop the chirality and operator indices for the remainder of this section.  
While the normalization of the form factors  requires a lattice-QCD calculation, see Table~\ref{tab:lattice}, their momentum dependence is strongly constrained by unitarity, in terms of the imaginary part that is generated by the meson--nucleon rescattering, see Fig.~\ref{fig:unitarity}. 

We write the corresponding amplitude for $N(p) P(k)\to N(p') P(k')$ as
\beq
T_{PN}=\bar u_N(p')\bigg[A(s,t)+\frac{\slashed{k}+\slashed{k'}}{2}B(s,t)\bigg]u_N(p),
\eeq
with Mandelstam variables 
\beq
s=(p+k)^2,\qquad t=(p-p')^2,
\eeq
where the latter is related to the scattering angle $z$ via
\beq
t=-2|\kvec|^2(1-z),\qquad |\kvec|^2=\frac{\lambda\big(s,\mN^2,\MP^2\big)}{4s},
\eeq
and $\lambda(a,b,c)=a^2+b^2+c^2-2(ab+ac+bc)$. The partial waves are obtained from the projection~\cite{Frazer:1960zz}
\begin{align}
 f_{l\pm}(s)&=\frac{1}{16\pi \sqrt{s}}\bigg\{(E+\mN)\big[A_l(s)+(\sqrt{s}-\mN)B_l(s)\big]\notag\\
 &+(E-\mN)\big[-A_{l\pm 1}(s)+(\sqrt{s}+\mN)B_{l\pm 1}(s)\big]\bigg\},
\end{align}
where $E=(s+\mN^2-\MP^2)/(2\sqrt{s})$, $l$ is the orbital angular momentum,  $j=|l\pm 1/2|$, and
\beq
Y_l(s)=\int_{-1}^1dz\,P_l(z)Y(s,t)\Big|_{t=-2|\kvec|^2(1-z)},
\eeq
with $Y\in\{A,B\}$ and Legendre polynomials $P_l(z)$. 

The unitarity constraint from the rescattering diagram 
in Fig.~\ref{fig:unitarity} becomes shortest once expressed in terms of the linear combinations
\beq
\label{Wpm}
W_\pm(s)=W_0(s)\pm \frac{\sqrt{s}}{\mN} W_1(s),
\eeq
both of which fulfill unitarity relations\footnote{These relations
are obtained by evaluating the unitarity diagram in Fig.~\ref{fig:unitarity}. Since $W_\pm$ are unitarized with different $\pi N$ phase shifts, the result depends critically on the sign convention for $W_1$; we follow the choice of Refs.~\cite{JLQCD:1999dld,Yoo:2021gql}.} 
\begin{align}
\label{unitarityWpm}
 \Im W_+(s)&=|\kvec| W_+(s) \big[f_{0+}(s)\big]^*,\notag\\
 \Im W_-(s)&=|\kvec| W_-(s) \big[f_{1-}(s)\big]^*,
\end{align}
that can be solved explicitly in terms of an Omn\`es representation~\cite{Omnes:1958hv}. 
To this end, one writes the partial waves in terms of phase shifts $\delta_{l\pm}^I$,
\beq
f_{l\pm}^I(s)=\frac{e^{i\delta_{l\pm}^I}}{|\kvec|}\sin\delta_{l\pm}^I,
\eeq
where $I$ refers to the isospin of the $PN$ system. Diagonalizing the unitarity relation in isospin space, we obtain that the solutions for $U_{1A}$ and $S_{5A}$ have $I=1/2$, $S_{1A}$ and $S_{2A}+S_{4A}$ have $I=1$, and $S_{3A}$ and $S_{2A}-S_{4A}$ have $I=0$, see~\ref{app:isospin} for details. These isospin numbers match the baryon poles ($B=N,\Sigma,\Lambda$, respectively) that can be present in the matrix elements. The ChPT analysis from Ref.~\cite{JLQCD:1999dld} shows that such poles only occur in $W_-$, with residues
\beq
\label{res}
\text{Res}\, W_-(s)\big|_{s=m_B^2}=\frac{2m_B^3}{\mN}W_1(0). 
\eeq
The rescattering is described by Omn\`es factors
\beq
\Omega^I_{l\pm}(s)=\exp\Bigg\{\frac{s}{\pi}\int_{s_\text{th}}^\infty ds'\frac{\delta_{l\pm}^I(s')}{s'(s'-s)}\Bigg\},
\eeq
where $s_\text{th}=(\mN+M_P)^2$, and if it were not for the kinematic singularity due to $\sqrt{s}$ in Eq.~\eqref{Wpm}, the solution for $W_\pm(s)$ would simply become a polynomial times $\Omega^I_{l\pm}(s)$. Such kinematic singularities often arise in the transition from a set of scalar functions for which dispersion relations can be written, $W_{0,1}(s)$, to the one 
for which unitarity constraints are simplest, $W_\pm(s)$, see Refs.~\cite{Colangelo:2014dfa,Colangelo:2015ama,Colangelo:2017qdm,Colangelo:2017fiz,Hoferichter:2019nlq,Hoferichter:2020lap}  for extensive discussions of this point in the context of hadronic light-by-light scattering. In this case, a solution can be found by diagonalizing the kernel functions in the dispersion integrals~\cite{Hoferichter:2019nlq}, but here this strategy fails because $W_\pm$ are unitarized with different phase shifts, in such a way that the linear combinations of Eq.~\eqref{unitarityWpm} that reproduce $W_{0,1}$ no longer take the form of a simple Omn\`es problem. Instead, we construct a solution as follows: first, we observe that in the linear combination $W_0(s)=(W_+(s)+W_-(s))/2$ the square-root singularity drops out. Since $\Omega_\pm^I$ takes care of the right-hand cut, we can therefore make the ansatz\footnote{Unitarity and analyticity determine the Omn\`es solutions up to a polynomial, whose degree can often be constrained by the asymptotic behavior of the amplitude or form factor, see, e.g., Ref.~\cite{Colangelo:2018mtw} for the case of the electromagnetic form factor of the pion. Given the limited kinematic range accessible here, we instead construct the solutions with the minimal amount of polynomial freedom required to satisfy all available constraints.} 
\beq
\label{W0}
W_0(s)=W_0(0)\Big[(1-\alpha)\Omega_{0+}^I(s)+\alpha \frac{m_B^2}{m_B^2-s}\Omega_{1-}^I(s)\Big],
\eeq
where we added the single-baryon pole that originates from $W_-(s)$. Next, we observe that the product
\beq
\label{W1}
W_+(s)W_-(s)=\big[W_0(s)\big]^2-\frac{s}{\mN^2}\big[W_1(s)\big]^2
\eeq
is again free of kinematic singularities, and fulfills an Omn\`es representation using the sum of the two phase shifts. In this case, we can thus make the ansatz
\beq
\label{WpWm}
W_+(s)W_-(s)=\big[W_0(0)\big]^2\Omega_{0+}^I(s)\Omega_{1-}^I(s)\frac{m_B^2}{m_B^2-s}(1+\beta s),
\eeq
where we kept a non-trivial order in the polynomial to be able to impose the normalization of $W_1(s)$. This yields 
\beq
\beta = \big(1-2\alpha\big)\bigg[\dot\Omega_{0+}^I-\dot\Omega_{1-}^I-\frac{1}{m_B^2}\bigg]-\frac{\big[W_1(0)\big]^2}{\mN^2 \big[W_0(0)\big]^2},
\eeq
where $\dot\Omega_{l\pm}^I=d\Omega_{l\pm}^I/ds|_{s=0}$. Finally, we determine $\alpha$ from the residue~\eqref{res}. Combining $W_0(s)$ from Eq.~\eqref{W0} and $W_+(s)W_-(s)$ from Eq.~\eqref{WpWm} gives $W_1(s)$ via Eq.~\eqref{W1}, where the sign needs to be chosen in accordance with the overall signs determined in lattice QCD, see Table~\ref{tab:lattice}. In particular, $W_\pm(s)$ can be identified by the fact that the single-baryon pole only contributes to $W_-$. Matching its residue to Eq.~\eqref{res} gives
\beq
\alpha=-\frac{m_B}{\mN}\frac{W_1(0)}{W_0(0)},
\eeq
which concludes the derivation of our dispersive representation for $W_{0,1}(s)$. The resulting expression implements the normalizations from lattice QCD, fulfills the unitarity constraint~\eqref{unitarityWpm}, and in the limit $\Omega^I_{l\pm}(s)=1$ reduces to the tree-level ChPT result~\cite{JLQCD:1999dld}
\begin{align}
\label{ChPT}
W_0^\text{ChPT}(s)&=W_0(0)\bigg[1-\frac{m_B}{\mN}\frac{W_1(0)}{W_0(0)}\frac{s}{m_B^2-s}\bigg],\notag\\
W_1^\text{ChPT}(s)&=W_1(0)\frac{m_B^2}{m_B^2-s},
\end{align}
in such a way that the dispersive result can be considered a unitarized version of Eq.~\eqref{ChPT}.

\begin{figure}[t]
\centering
 \includegraphics[width=0.49\linewidth]{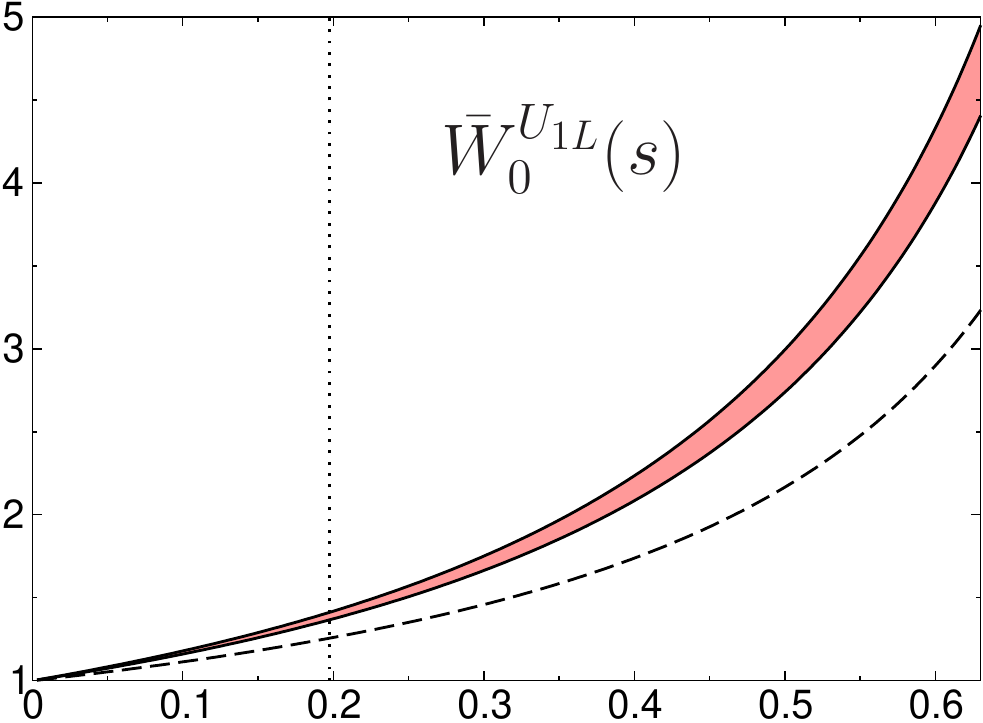}
 \includegraphics[width=0.49\linewidth]{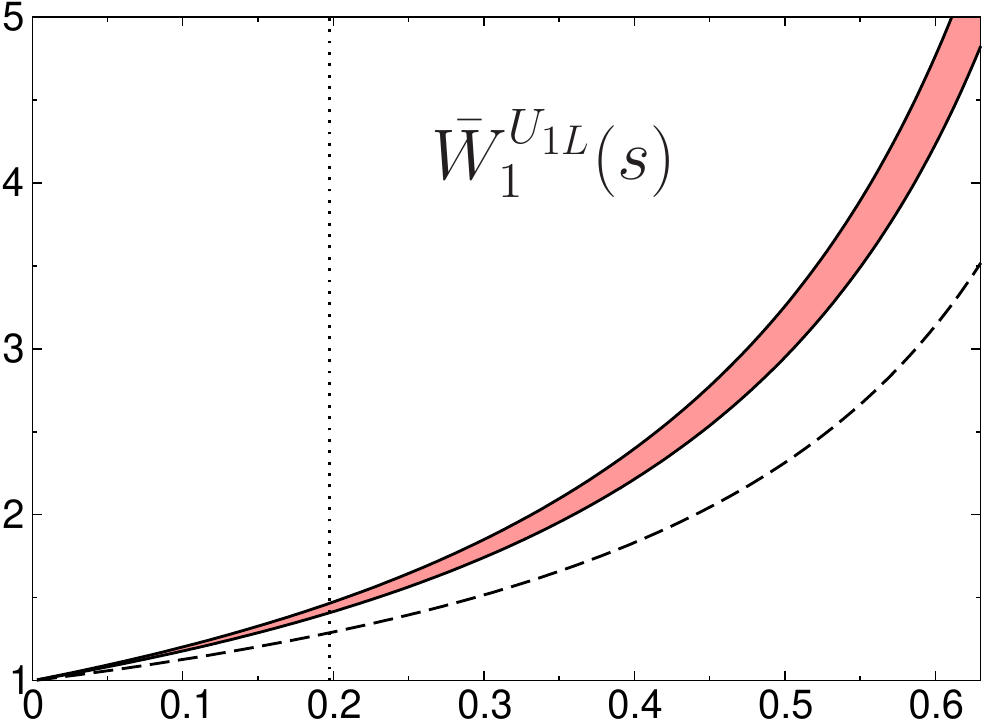}\\
 \includegraphics[width=0.49\linewidth]{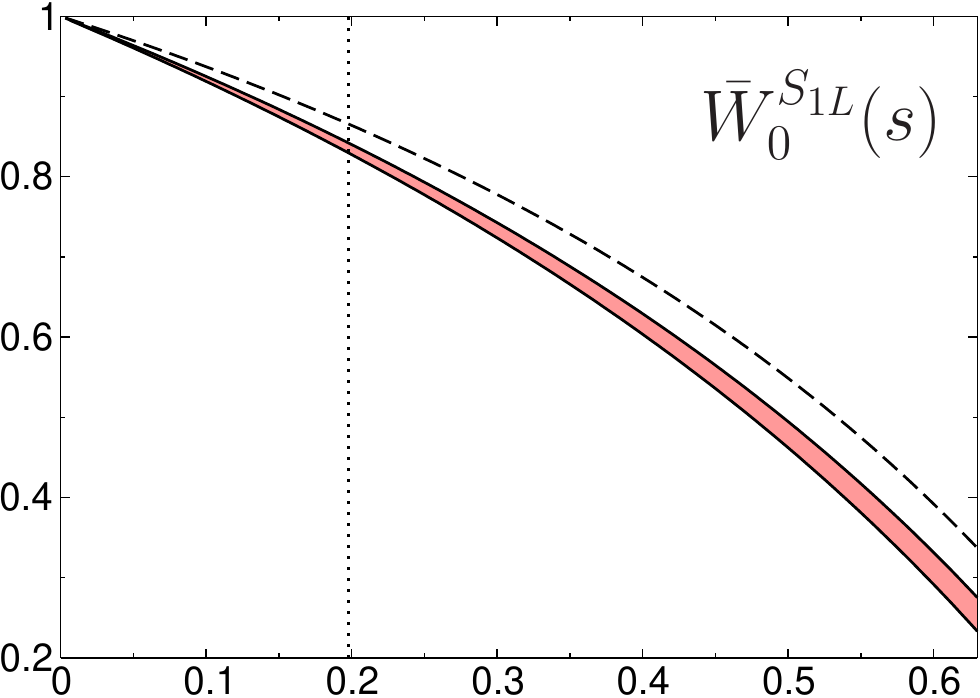}
 \includegraphics[width=0.49\linewidth]{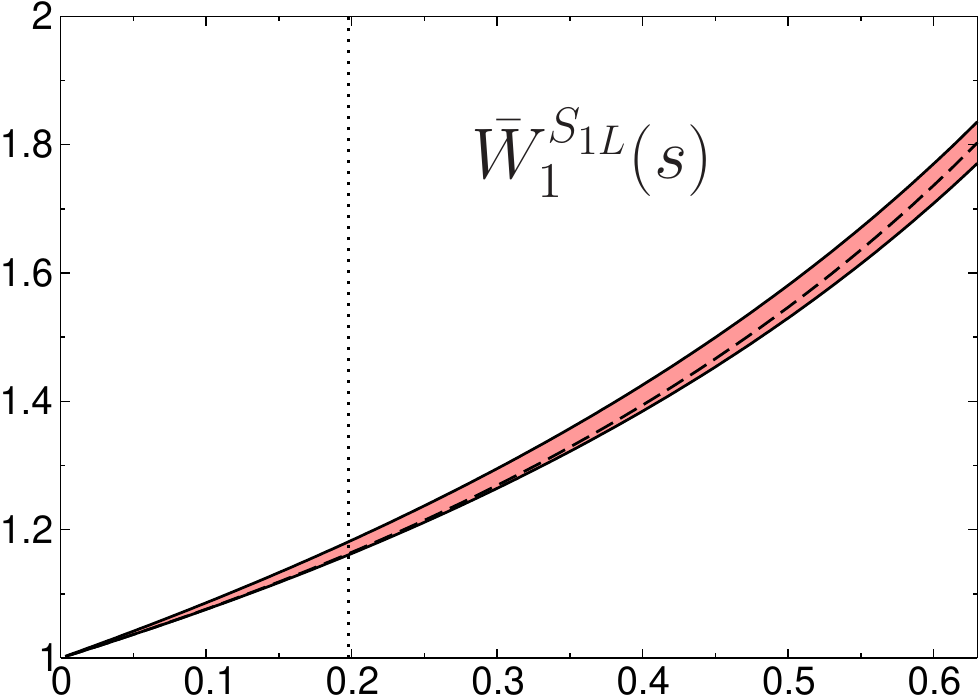}\\
 \includegraphics[width=0.49\linewidth]{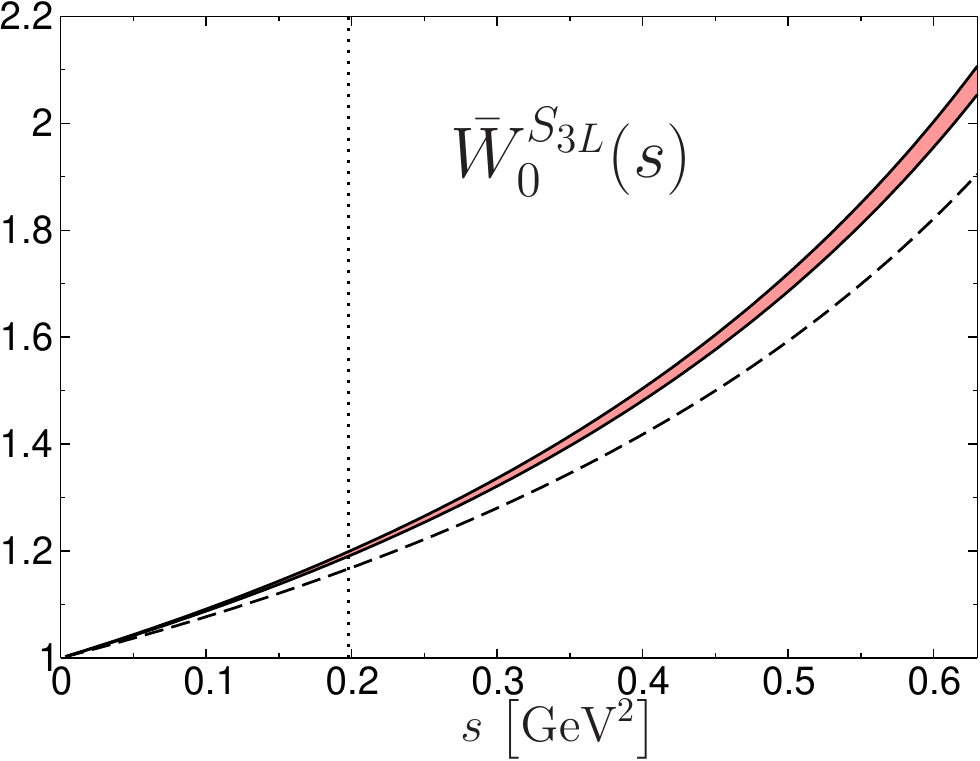}
 \includegraphics[width=0.49\linewidth]{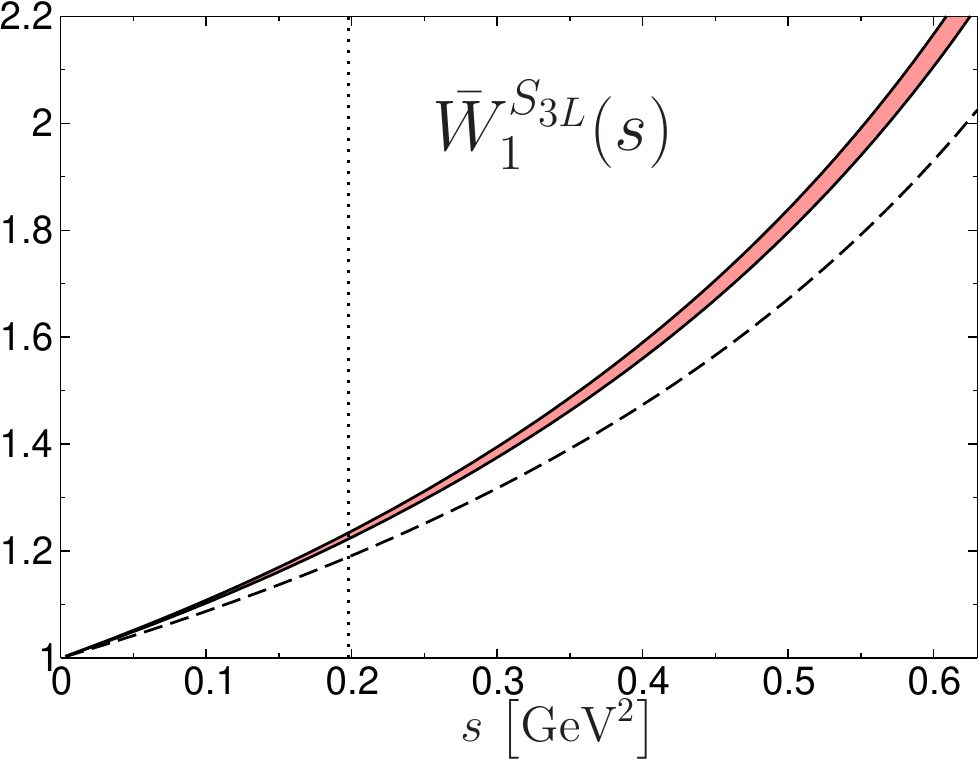}
 \caption{Form factors $\bar W_i^{X_{iL}}(s)=W_i^{X_{iL}}(s)/W_i^{X_{iL}}(0)$ for representative cases involving $\pi N$, $I=1/2$ (top), $KN$, $I=1$ (middle), and $KN$, $I=0$ (bottom) rescattering. The dashed line indicates the result without these unitarity corrections, which corresponds to the tree-level ChPT result~\eqref{ChPT}. The results are shown for the entire range $s\leq (\mN-M_\pi)^2$ relevant for $p\to\pi^0\ell^+\bar\nu_\ell\nu_\tau$, while the dotted line indicates the kinematic limit $s=(\mN-M_K)^2$ for the kaon modes.}
 \label{fig:formfactors}
\end{figure}

Our numerical results are shown in Fig.~\ref{fig:formfactors} for a representative set of form factors. The low-energy phase shifts for the $\pi N$ case are known to high precision from Roy--Steiner equations~\cite{Ditsche:2012fv,Hoferichter:2015hva,Hoferichter:2015tha,Siemens:2016jwj}, in combination with experimental input from pionic atoms~\cite{Strauch:2010vu,Hennebach:2014lsa,Hirtl:2021zqf,Baru:2010xn,Baru:2011bw,Hoferichter:2023ptl} or cross sections~\cite{RuizdeElvira:2017stg}, while modern phase-shift analyses extend to $2.5\GeV$~\cite{Arndt:2006bf,Workman:2012hx}. Since the main uncertainties arise from the high-energy continuation of the phase shift in the evaluation of $\Omega^I_{l\pm}$---compared to which other uncertainties, e.g., from higher intermediate states 
or higher orders in the matching to ChPT will be subdominant---we use the latter solution, and estimate the uncertainties from the variation observed when keeping the phase constant above a matching point chosen within $[1.8,2.5]\GeV$ in the evaluation of the integral in the Omn\`es factors.   
Proceeding similarly for the $KN$ case~\cite{Hyslop:1992cs,Arndt:2003xz} leads to the bands in Fig.~\ref{fig:formfactors}, in comparison to the dashed lines representing the tree-level ChPT result~\eqref{ChPT}. As expected, one sees that the unitarity corrections are most relevant for the decay into a pion, because the phase space in the kaon modes is much more limited.

\section{Tau-induced proton decay}

The rate for the two-body decay $N\to P\bar \ell$ can be written as~\cite{Yoo:2021gql}
\beq
\Gamma[N\to P\bar\ell]=\frac{|\qvec| E_\ell}{8\pi\mN}\Big[\sum_{A=L,R}\big|W_A^\ell\big|^2-\frac{2m_\ell}{E_\ell}\text{Re}\big(W_L^\ell W_R^{\ell*}\big)\Big],
\eeq
where the form factors
\begin{align}
\label{WLR}
W_L^\ell(s)&=\sum_{\alpha,A}\bigg[C^{AL}_\alpha W_0^{X^{AL}}(s)-\frac{s}{\mN m_\ell}C^{AR}_\alpha W_1^{X^{AR}}(s)\bigg],\notag\\
W_R^\ell(s)&=\sum_{\alpha,A}\bigg[C^{AR}_\alpha W_0^{X^{AR}}(s)-\frac{s}{\mN m_\ell}C^{AL}_\alpha W_1^{X^{AL}}(s)\bigg],
\end{align}
are evaluated at $s=m_\ell^2$ and
\beq
E_\ell=\frac{\mN^2+m_\ell^2-M_P^2}{2\mN},\quad 
|\qvec|=\frac{\lambda^{1/2}(\mN^2,m_\ell^2,M_P^2)}{2\mN}.
\eeq
The sums in Eq.~\eqref{WLR} run over all contributing operators $\alpha$ and the two possible parity assignments $A=L,R$, see~\ref{app:twobody} for the explicit expressions in a given decay channel. Using the lifetime limits from Table~\ref{tab:lifetime}, stringent constraints $|C_\alpha|\lesssim (10^{-15}/\GeV)^{2}$ on most Wilson coefficients (given at the nucleon scale) can be obtained. An exception concerns those operators that only contribute to the processes with a charged lepton, since the decay with a $\tau$ final state is kinematically forbidden. Therefore, no limits on the corresponding $\tau$-flavored Wilson coefficients can be extracted from two-body decays.  

\begin{figure}[t]
\centering
 \includegraphics[width=0.9\linewidth]{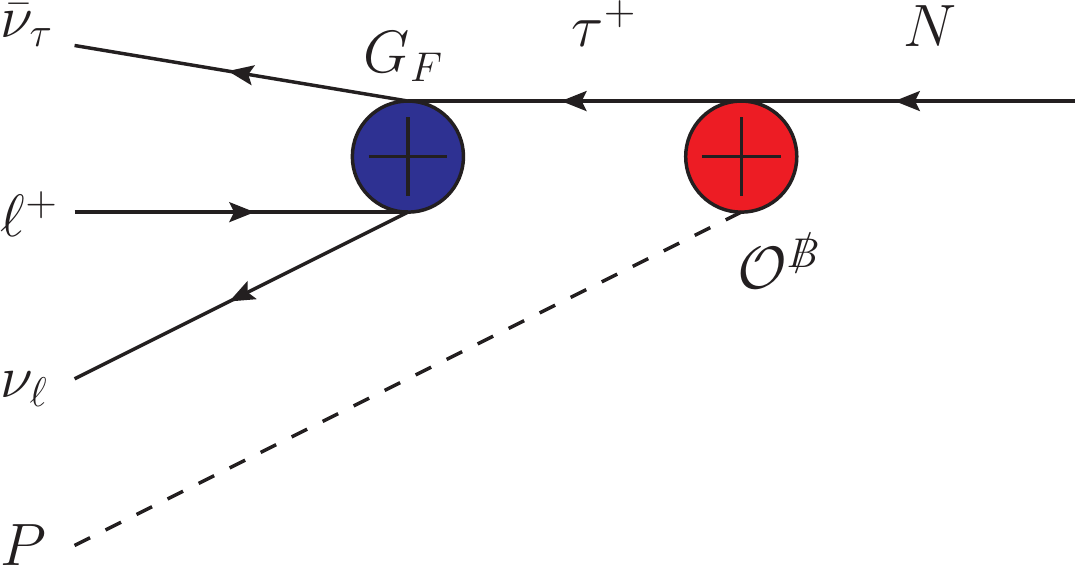}
 \caption{Decay $N\to P \ell^+\nu_\ell\bar\nu_\tau$ via an off-shell $\tau$. ${\mathcal O}^{\slashed{B}}$ and $G_F$ indicate the insertion of the $B$-violating operator and the electroweak vertex, respectively. }
 \label{fig:4body}
\end{figure}

However, the $\tau$ lepton can be off-shell, thus inducing four-body decays, see Fig.~\ref{fig:4body}, whose width can be expressed as
\begin{align}
\label{4body}
 \Gamma[N\to P \ell^+\nu_\ell\bar\nu_\tau]&=
 \int_{0}^{s_\text{max}}ds \frac{G_F^2s^2\lambda^{1/2}(\mN^2,s,M_P^2)}{6144\pi^5\mN^3(m_\tau^2-s)^2}\notag\\
 &\hspace{-11pt}\times \bigg[\big(\mN^2+s-M_P^2\big)\Big(m_\tau^2
 |W_R^\tau|^2+s |W_L^\tau|^2\Big)\notag\\
 &-4s\mN m_\tau \Re(W_L^\tau W_R^{\tau *})\bigg],
\end{align}
where $s_\text{max}=(\mN-M_P)^2$. 
For simplicity, Eq.~\eqref{4body} is given in the limit $m_\ell=0$, see~\ref{app:fourbody} for the full expression. 
Numerically, we find for the special case $C^{AB}_{\tau \alpha}\equiv C^{LL}_{\tau \alpha}=C^{LR}_{\tau \alpha}=C^{RL}_{\tau \alpha}=C^{RR}_{\tau \alpha}$
\begin{align}
\Gamma[p\to \pi^0 \ell^+\nu_\ell\bar\nu_\tau]&= 14.3(1.8)\bigg|\frac{C^{AB}_{\tau udu}}{(10^{10}\GeV)^{-2}}\bigg|^2\frac{1}{10^{30}\text{yr}},\notag\\
\Gamma[n\to \pi^- \ell^+\nu_\ell\bar\nu_\tau]&= 26.2(3.3)\bigg|\frac{C^{AB}_{\tau udu}}{(10^{10}\GeV)^{-2}}\bigg|^2\frac{1}{10^{30}\text{yr}},\notag\\
\Gamma[p\to K^0 \ell^+\nu_\ell\bar\nu_\tau]&= 0.81(1)\bigg|\frac{C^{AB}_{\tau usu}}{(10^{10}\GeV)^{-2}}\bigg|^2\frac{1}{10^{30}\text{yr}},
\end{align}
where the uncertainty corresponds to the bands in Fig.~\ref{fig:formfactors} (the coefficients without rescattering corrections are $7.2$, $13.3$, and $0.84$, respectively). 
As expected, the resulting numerical limits are significantly weaker than in the two-body decay, $|C_\alpha|\lesssim (10^{-10}/\GeV)^{2}$ at the nucleon scale---due to a combination of the weaker inclusive limits, see Table~\ref{tab:lifetime}, and the suppression from $G_F$ as well as the four-particle phase space---but we emphasize that the corresponding Wilson coefficients were previously completely unconstrained. 
A comprehensive analysis in SMEFT, including the pertinent RG corrections~\cite{Alonso:2014zka}, will be given in Ref.~\cite{RG}.     

\section{Conclusions}
 
Processes such as $p\to \pi^0 \ell^+\nu_\ell\bar\nu_\tau$ are phenomenologically interesting, as they can probe Wilson coefficients that are not accessible in simpler two-body decays. I.e., while the decay to $\tau$ leptons is kinematically forbidden, $B$-violating operators involving a $\tau$ lepton do become constrained by such four-body processes. However, for calculating the corresponding decay widths, the momentum dependence of the related hadronic matrix elements is necessary.

In this work, we examined these form factors for nucleon-to-meson transitions in dispersion theory and derived a representation that allows one to calculate the momentum dependence of the corresponding form factors including unitarity corrections from the meson--nucleon rescattering. With normalizations determined from lattice QCD, our results allowed us to calculate semileptonic nucleon decays involving multiple leptons in the final state~\cite{Faroughy:2014tfa,Hambye:2017qix,Dorsner:2022twk} and to derive novel bounds on the Wilson coefficients of previously unconstrained dimension-$6$ operators involving $\tau$ leptons.

\section*{Acknowledgments}

We thank Bastian Kubis and Peter Stoffer for helpful discussions, and Taku Izubuchi and Sergey Syritsyn for communication regarding Ref.~\cite{Yoo:2021gql}. 
Financial support from  the SNSF (Project Nos.\ PP00P21\_76884 and  PCEFP2\_181117) is gratefully acknowledged.

\appendix

\section{Operators in SMEFT}
\label{sec:SMEFT}

The four SU(2)$_L$-invariant dimension-$6$ operators giving rise to nucleon decay are
\begin{align}
\label{SMEFT}
Q_{duq} &= {\varepsilon ^{\alpha \beta \gamma }}{\varepsilon _{jk}}\left[ {{{\left( {d_p^\alpha } \right)}^T}Cu_r^\beta } \right]\left[ {{{\left( {q_s^{\gamma j}} \right)}^T}CL_t^k} \right],\notag\\
Q_{qqu} &= {\varepsilon ^{\alpha \beta \gamma }}{\varepsilon _{jk}}\left[ {{{\left( {q_p^{\alpha j}} \right)}^T}Cq_r^{\beta k}} \right]\left[ {{{\left( {u_s^\gamma } \right)}^T}C{e_t}} \right],\notag\\
Q_{qqq} &= {\varepsilon ^{\alpha \beta \gamma }}{\varepsilon _{jn}}{\varepsilon _{km}}\left[ {{{\left( {q_p^{\alpha j}} \right)}^T}Cq_r^{\beta k}} \right]\left[ {{{\left( {q_s^{\gamma m}} \right)}^T}CL_t^n} \right],\notag\\
Q_{duu} &= {\varepsilon ^{\alpha \beta \gamma }}\left[ {{{\left( {d_p^\alpha } \right)}^T}Cu_r^\beta } \right]\left[ {{{\left( {u_s^\gamma } \right)}^T}C{e_t}} \right],
\end{align}
where $q$/$L$ denote the quark/lepton SU(2)$_L$ doublets, $u$, $d$, and $e$ the quark and lepton singlets, $p, \ldots$/$j, \ldots$/$\alpha,\ldots$ are generation/SU(2)$_L$/color indices, and $C$ denotes the charge conjugation operator.  After electroweak symmetry breaking, we can write the resulting effective Lagrangian below the weak scale as
\begin{align}
 \Lagr&=\bar\ell^c_t\Big[ -C_{qqq}^{111t}Q_{udu}^{LL} 
 +2C_{qqu}^{111t}Q_{udu}^{LR}
 -C_{duq}^{111t}Q_{udu}^{RL}\notag\\
 &\quad-C_{duu}^{111t}Q_{udu}^{RR} 
 -C_{qqq}^{121t}Q_{usu}^{LL}
+\big(C_{qqu}^{121t}+C_{qqu}^{211t}\big)Q_{usu}^{LR}\notag\\
 &\quad-C_{duq}^{211t}Q_{usu}^{RL}
 -C_{duu}^{211t}Q_{usu}^{RR}
 \Big]\notag\\
 &+\bar\nu^c_t\Big[C_{qqq}^{111t} Q_{udd}^{LL}+C_{duq}^{111t} Q_{udd}^{RL}
 +C_{qqq}^{211t} Q_{usd}^{LL}\notag\\
 &\quad+C_{duq}^{211t} Q_{usd}^{RL}
 +\big(C_{qqq}^{121t}-C_{qqq}^{211t}\big)Q_{dsu}^{LL}\notag\\
 &\quad+C_{qqq}^{112t}Q_{uds}^{LL}
 +C_{duq}^{112t}Q_{uds}^{RL}
 \Big],
\end{align}
where the Wilson coefficients refer to the respective flavor combinations in Eq.~\eqref{SMEFT}, the quark part of the effective operators is defined in Eq.~\eqref{QABi}, and of course RG corrections need to be added to relate low and high scale. Note that assuming that $B$ violation is realized far above the electroweak scale reduces the number of operators, otherwise, also different flavor combinations with $\Delta B\neq \Delta L$ appear~\cite{Kobach:2016ami,Jenkins:2017jig}. 

\section{Isospin symmetry}
\label{app:isospin}

In addition to the isospin relations given in the main text, isospin symmetry also determines the neutron matrix elements~\cite{Aoki:2013yxa}
\begin{align}
\label{isospin_neutron}
 \langle \pi^-|\left[\bar u^c P_A d\right]u_L|n\rangle&=-\sqrt{2} \langle \pi^0|\left[\bar u^c P_A d\right]d_L|n\rangle= U_{1A},\notag\\
 \langle K^+|\left[\bar d^c P_A s\right]d_L|n\rangle&=-S_{1A},\notag\\ 
 \langle K^0|\left[\bar u^c P_A s\right]d_L|n\rangle&=-S_{4A},\notag\\
 \langle K^0|\left[\bar u^c P_A d\right]s_L|n\rangle&= S_{3A},\notag\\
  \langle K^0|\left[\bar d^c P_A s\right]u_L|n\rangle&= -S_{2A},\notag\\
  \langle \eta|\left[\bar u^c P_A d\right]d_L|n\rangle&= S_{5A}.
\end{align}
To work out the isospin quantum numbers of the rescattering amplitudes, we sum up all possible intermediate states in each charge channel and use the isospin relations~\eqref{isospin_neutron} to obtain a closed system of unitarity relations
\begin{align}
 \Im U_{1A}&=\Big[\langle\pi^0p|T|\pi^0 p\rangle-\sqrt{2}\langle\pi^0 p|T|\pi^+ n\rangle\Big]^* U_{1A},\notag\\
 \Im U_{1A}&=\Big[\langle\pi^-p|T|\pi^- p\rangle-\frac{1}{\sqrt{2}}\langle\pi^- p|T|\pi^0 n\rangle\Big]^* U_{1A},\notag\\
 \Im S_{1A}&=\langle\bar K^0 p|T|\bar K^0 p\rangle^* S_{1A},\notag\\
 \Im S_{3A}&=\Big[\langle K^-p|T|K^- p\rangle+\langle K^- p|T|\bar K^0 n\rangle\Big]^* S_{3A},\notag\\
\text{Im}\!\begin{bmatrix}
     S_{2A}\\ S_{4A}
    \end{bmatrix}
&=\begin{bmatrix}
   \langle K^- p|T|K^- p\rangle & -\langle K^- p|T|\bar K^0 n\rangle\\
   -\langle K^- p|T|\bar K^0 n\rangle & \langle K^- p|T|K^- p\rangle
  \end{bmatrix}^*
\begin{bmatrix}
     S_{2A}\\ S_{4A}
    \end{bmatrix},\notag\\
 \Im S_{5A}&=\langle\eta p|T|\eta p\rangle^* S_{5A}.
\end{align}
Expressing each amplitude in terms of isospin quantum numbers and diagonalizing the relation for $S_{2A}$, $S_{4A}$, we obtain
\begin{align}
\Im U_{1A}&=\big[T^{1/2}\big]^* U_{1A},\notag\\
\Im S_{5A}&=\big[T^{1/2}\big]^* S_{5A},\notag\\
\Im S_{1A}&=\big[T^{1}\big]^* S_{1A},\notag\\
\Im\big(S_{2A}+S_{4A}\big)&=\big[T^1\big]^*\big(S_{2A}+S_{4A}\big),\notag\\
\Im S_{3A}&=\big[T^{0}\big]^* S_{3A},\notag\\
\Im\big(S_{2A}-S_{4A}\big)&=\big[T^0\big]^*\big(S_{2A}-S_{4A}\big),
\end{align}
which determines the isospin quantum numbers of the rescattering corrections as stated in the main text. Moreover, this analysis in terms of isospin symmetry leads one in a straightforward way to the relations~\eqref{isospinfierz}, since the corresponding matrix elements carry the same isospin. Further relations could be derived from SU(3) flavor symmetry, but would only become relevant in the study of hyperon decays.

\section{Two-body decay}
\label{app:twobody}

The explicit decomposition of $W_{A}$ in Eq.~\eqref{WLR} for the different decay channels in terms of SMEFT Wilson coefficients reads
 \begin{align}
  W_L[p\to\pi^0 \ell^+]&\simeq -\frac{1}{\sqrt{2}}\big(C_{qqq}^{111t}W_0^{U_{1L}}+C^{111t}_{duq}W_0^{U_{1R}}\big),\notag\\
  W_R[p\to\pi^0 \ell^+]&\simeq\frac{1}{\sqrt{2}}\big(2C_{qqu}^{111t}W_0^{U_{1R}}-C^{111t}_{duu}W_0^{U_{1L}}\big),\notag\\
 W_L[p\to\pi^+ \bar \nu]&\simeq C_{qqq}^{111t}W_0^{U_{1L}}+C^{111t}_{duq}W_0^{U_{1R}},\notag\\ 
 W_L[p\to K^0 \ell^+]&\simeq-C_{qqq}^{121t}W_0^{S_{1L}}-C^{211t}_{duq}W_0^{S_{1R}},\notag\\
 W_R[p\to K^0 \ell^+]&\simeq\big(C_{qqu}^{121t}+C_{qqu}^{211t}\big)W_0^{S_{1R}}-C^{211t}_{duu}W_0^{S_{1L}},\notag\\
 W_L[p\to K^+ \bar \nu]&\simeq\big(C_{qqq}^{112t}+C_{qqq}^{211t}\big)W_0^{S_{3L}}+C_{qqq}^{121t}W_0^{S_{4L}}\notag\\
 &+C^{211t}_{duq}W_0^{S_{2R}}+C_{duq}^{112t}W_0^{S_{3R}}
 ,\notag\\ 
 W_L[p\to\eta \ell^+]&\simeq-C_{qqq}^{111t}W_0^{S_{5L}}-C^{111t}_{duq}W_0^{S_{5R}},\notag\\
 W_R[p\to\eta \ell^+]&\simeq2C_{qqu}^{111t}W_0^{S_{5R}}-C^{111t}_{duu}W_0^{S_{5L}},\notag\\
 W_L[n\to\pi^- \ell^+]&\simeq-C_{qqq}^{111t}W_0^{U_{1L}}-C^{111t}_{duq}W_0^{U_{1R}},\notag\\
 W_R[n\to\pi^- \ell^+]&\simeq2C_{qqu}^{111t}W_0^{U_{1R}}-C^{111t}_{duu}W_0^{U_{1L}},\notag\\
 W_L[n\to\pi^0 \bar \nu]&\simeq-\frac{1}{\sqrt{2}}\big(C_{qqq}^{111t}W_0^{U_{1L}}+C^{111t}_{duq}W_0^{U_{1R}}\big),\notag\\
  W_L[n\to K^0 \bar \nu]&\simeq\big(C_{qqq}^{112t}+C_{qqq}^{211t}\big)W_0^{S_{3L}}-C_{qqq}^{121t}W_0^{S_{2L}}\notag\\
 &-C^{211t}_{duq}W_0^{S_{4R}}+C_{duq}^{112t}W_0^{S_{3R}}
 ,\notag\\
 W_L[n\to\eta \bar\nu]&\simeq C_{qqq}^{111t}W_0^{S_{5L}}+C^{111t}_{duq}W_0^{S_{5R}},
 \end{align} 
where the dependence of $W_{0}$ on $s$ has been suppressed and we only give the $W_0$ term since the correction from $W_1$ follows directly from Eq.~\eqref{WLR}. Equation~\eqref{isospinfierz} has been used to simplify the expressions for $W_L[p\to K^+ \bar \nu]$ and $W_L[n\to K^0 \bar \nu]$. The notation using $\simeq$ is meant to indicate that the expression in terms of SMEFT coefficients does not apply at the low scale without accounting for RG corrections.

\section{Four-body decay}
\label{app:fourbody}

Denoting momenta as 
$p(p)\to \tau^+(q)P(p')$, $\tau^+(q)\to \bar \nu_\tau(q_1)\ell^+(q_2)\nu_\ell(q_3)$, we find as intermediate result for the spin-averaged squared matrix element:
\begin{align}
 |\bar M|^2&=\frac{64G_F^2q_1\cdot q_2}{(m_\tau^2-q^2)^2}\bigg[\Big(m_\tau^2 |W_R^\tau|^2-q^2 |W_L^\tau|^2\Big)p\cdot q_3\notag\\
 &+2\Big(|W_L^\tau|^2p\cdot q-\mN m_\tau \Re(W_L^{\tau} W_R^{\tau*})\Big)q\cdot q_3\bigg].
\end{align}
We perform the phase-space integration following Refs.~\cite{Guo:2011ir,Zanke:2021wiq} by subsequent reduction to two-body decays. This gives
\begin{align}
 \Gamma[N\to P \ell^+\nu_\ell\bar\nu_\tau]&=
 \frac{G_F^2}{512\pi^5\mN^3}\int_{m_\ell^2}^{s_\text{max}}\hspace{-1pt}ds\frac{\lambda^{1/2}(\mN^2,s,M_P^2)}{s^2(m_\tau^2-s)^2}\notag\\
 &\hspace{-11pt}\times \bigg[\big(\mN^2+s-M_P^2\big)\Big(m_\tau^2
 |W_R^\tau|^2+s |W_L^\tau|^2\Big)\notag\\
 &-4s\mN m_\tau \Re(W_L^{\tau} W_R^{\tau *})\bigg]I_\ell(s),
\end{align}
where
\begin{align}
I_\ell(s)&=\int_{m_\ell^2}^s dk^2 \frac{(k^2-m_\ell^2)^2(s-k^2)^2}{k^2}\\
&=\frac{1}{12}\Big(s^4-8s^3 m_\ell^2+8s m_\ell^6-m_\ell^8\Big)+s^2 m_\ell^4\log\frac{s}{m_\ell^2}\notag
\end{align}
coincides with the standard phase-space factor for the leptonic decay of the muon, see, e.g., Refs.~\cite{Sirlin:2012mh,Ferroglia:2013dga}.

\bibliographystyle{apsrev4-1_mod}
\balance
\biboptions{sort&compress}
\bibliography{bib}

\end{document}